\documentclass[twocolumn]{IEEEtran}
\usepackage[usenames,dvipsnames]{color}
\usepackage{subfigure}
\usepackage{graphicx}
\usepackage{amsmath}
\usepackage{color}
\usepackage{multicol}
\usepackage{amssymb}

\usepackage{amsfonts}%
\usepackage{verbatim}%
\usepackage{enumerate}%
\usepackage{psfrag}
\usepackage{epsfig}
\usepackage{multicol}
\usepackage{setspace}

\usepackage{algorithm}
\usepackage{algorithmic}
\usepackage{cite}
\usepackage{float}
\begin{document}

\title{Optimal Selection of Spectrum Sensing Duration for an Energy Harvesting Cognitive Radio}
\author{ Ahmed El Shafie$^\dagger$, Ahmed Sultan$^*$\\
\small \begin{tabular}{c}
$^\dagger$Wireless Intelligent Networks Center (WINC), Nile University, Giza, Egypt. \\
$^*$Department of Electrical Engineering, Alexandria University, Alexandria, Egypt. \\
\end{tabular}
}
\date{}
\maketitle
\thispagestyle{empty}
\pagestyle{empty}
\begin{abstract}
In this paper, we consider a time-slotted cognitive radio (CR) setting with buffered and energy harvesting primary and CR users. At the beginning of each time slot, the CR user probabilistically chooses the spectrum sensing duration from a predefined set. If the primary user (PU) is sensed to be inactive, the CR user accesses the channel immediately. The CR user optimizes the sensing duration probabilities in order to maximize its mean data service rate with constraints on the stability of the primary and cognitive queues. The optimization problem is split into two subproblems. The first is a linear-fractional program, and the other is a linear program. Both subproblems can be solved efficiently.
\end{abstract}
\begin{IEEEkeywords}
Cognitive radio, energy harvesting, queue stability, dominant system, linear-fractional programming.
\end{IEEEkeywords}
\vspace{-0.1cm}
\section{Introduction}
\vspace{-0.1cm}
Cognitive radio (CR) technology has been proposed as a solution to the problem of spectrum scarcity. When primary and cognitive users in a network are energy harvesters, proper energy management becomes crucial to network operation, especially that the network has a priority structure.

Data transmission by an energy harvesting transmitter with a rechargeable battery has received a lot of attention recently  \cite{lei2009generic,sharma2010optimal,ho2010optimal,hoang2009opportunistic,park2011energy,krikidis2012stability,pappas2012optimal,Sultan,Sult1210:Optimal}. The authors of \cite{lei2009generic} used a dynamic programming framework to derive the optimal online policy for controlling admissions into the data buffer.
Optimal energy management has been addressed in many
papers such as \cite{sharma2010optimal,ho2010optimal}. Sharma {\it et al.} \cite{sharma2010optimal} obtained throughput optimal
energy management policies for an energy harvesting
sensor node. The discounted throughput is maximized over an infinite horizon.
Throughput maximization via energy allocation over a finite horizon, taking into account a time varying channel and energy source, was investigated in \cite{ho2010optimal}.

In a cognitive setting, the authors of \cite{hoang2009opportunistic} investigate an energy constrained
cognitive terminal without explicitly involving an
energy queue. In \cite{park2011energy},
a Markov decision process (MDP) is adopted to obtain the
optimal secondary access policy under perfect spectrum sensing. The authors of \cite{krikidis2012stability} investigated the impact of cooperation on the stable throughput of the source in a wireless three-node network topology (source-relay-destination) with energy harvesting
nodes and bursty data traffic and without channel state information (CSI) at the transmitters. Pappas {\it et al.} \cite{pappas2012optimal} investigated a
scenario with one rechargeable primary user (PU) and one CR user (secondary user) and characterized the maximum stable throughput region. In \cite{Sultan}, the author investigated the optimal cognitive sensing
and access policies for a CR user with an energy queue based on the MDP. In \cite{Sult1210:Optimal}, we investigated the optimal random spectrum sensing and accessing for an energy harvesting cognitive node. The maximum secondary stable throughput was characterized with and without primary feedback leveraging.

In this work, we investigate buffered primary and secondary nodes, each with an energy queue to store energy harvested from the environment. We consider spectrum sensing errors. Unlike most of the existent works, we do not approximate the energy queue using the M/D/1 approximation where one packet is expended from each energy queue at each time slot (e.g., see \cite{pappas2012optimal,Sult1210:Optimal} and the references therein). An increased spectrum sensing duration improves the reliability of detecting primary activity, but it degrades the throughput. Hence, the CR user optimizes its choice of the sensing duration via a probabilistic selection of one of the feasible durations. To the best of our knowledge, the analysis of such setting involving buffered energy harvesting terminals is reported in this paper for the first time.

This paper is organized as follows. In the following section, we provide the system model adopted in this paper. In Section \ref{sec3}, we provide the queues service processes. Stability analysis and problem formulation are presented in Section \ref{sec4}. We present some numerical results and conclude the paper in Section \ref{sec5}.
\vspace{-0.2cm}
\section{SYSTEM MODEL}
\label{sec2}
\begin{figure}[t]
\normalcolor
\center
  \includegraphics[width=0.8\columnwidth]{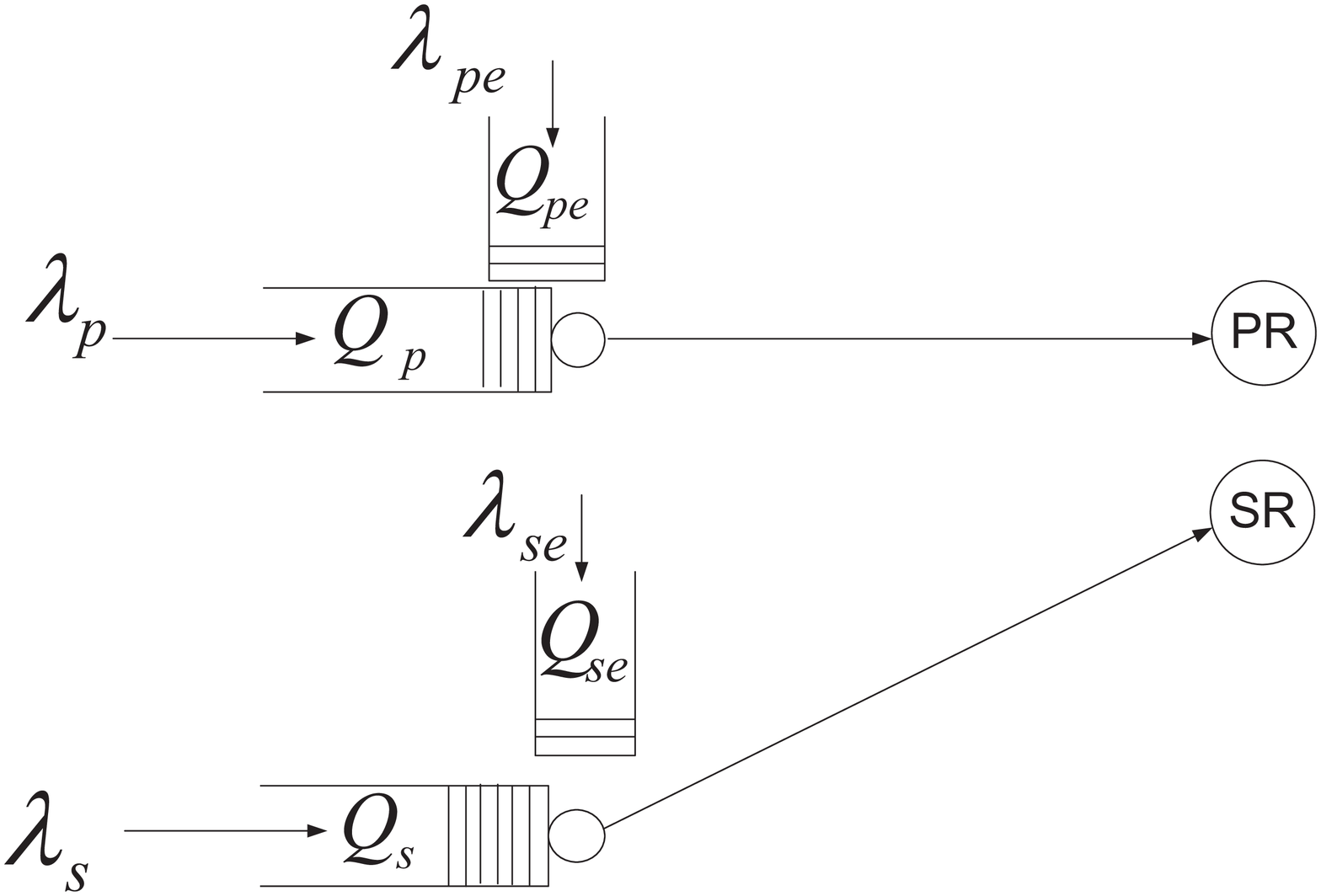}
  \caption{Primary and secondary links and queues. The solid lines are the communication channels. In the figure, the secondary receiver and the primary receiver are denoted by SR and PR, respectively.}\label{relayednw}
\end{figure}

\vspace{-0.2cm}
The network model consists of two different priority energy harvesters; one rechargeable PU and one rechargeable CR user as depicted in
Fig. \ref{relayednw}. The considered
setting can be viewed as a subsystem within a bigger network
with different primary and secondary pairs using orthogonal
frequency channels. In this paper, we focus on the analysis of the pair using
the same channel. The PU has unconditional access to the channel, and it starts the transmission at the beginning of the time slot. At the beginning of the time slot, the CR user decides the width of channel sensing duration from a predefined set of widths. It senses the channel for $\tau$ seconds from the beginning of the time slot. If the channel is sensed to be idle, the CR user accesses the channel with probability $1$. We consider a wireless collision channel, shared by the PU and the CR user. That is, all concurrent transmissions are assumed lost packets.

We assume that the primary transmitter has two queues (buffers); a queue to store the incoming data packets, denoted by $Q_{\rm p}$, and an energy queue to store energy packets, denoted by $Q_{\rm pe}$. The CR user has two queues, $Q_{\rm s}$ to store the arrived data packets and an energy queue for harvesting the energy packets from the environment, denoted by $Q_{\rm se}$. We assume that all buffers are of
infinite length (a similar assumption is used in \cite{pappas2012optimal,krikidis2012stability} and the references therein). In practice, if these buffers are large enough compared to data and energy packet sizes, then this is a reasonable approximation \cite{krikidis2012stability}. We consider time-slotted transmissions where all packets have
the same size, namely $b$ bits per packet, and each packet is transmitted over one time slot of duration $T$. We adopt a discrete-time late arrival model, which means that a newly arrived packet during a particular time slot cannot be transmitted during the slot itself even if the queue is empty. The packet arrival processes to the primary and secondary queues are Bernoulli processes. The mean arrival rate of $Q_k$, $k=\{\rm p,pe,s,se\}$, is  $\lambda_k\in[0,1]$ packets per time slot. At any time slot, the probability of having an arrival at $Q_k$ is $\lambda_k$.
The arrival processes are independent and identical random variables from time slot to time slot, from queue to queue and from terminal to terminal \cite{pappas2012optimal,krikidis2012stability,sadek}. The
Bernoulli energy arrival model is simple, but it captures the
random and sporadic availability of ambient energy sources \cite{pappas2012optimal,Sultan,Sult1210:Optimal,lei2009generic,medepally2009implications,michelusi2012optimal}.

 We do not assume CSI at the transmitting terminals. The channel gain, $h^{\rm t}_i$, of the link between any pair of
nodes is assumed to be constant during one slot, where $i$ reads `${\rm p}$' for the
link between the PU and its respective receiver and `${\rm s}$' for the link between the CR user and its respective receiver. The gain of link `$i$' is distributed according to a zero mean circularly symmetric
complex Gaussian random variable with variance $\sigma_{i}^2$, i.e., $\mathcal{CN}(0,\sigma_i^2)$, and independent for all $i$. Each receiver is affected by an additive white Gaussian noise (AWGN) with zero mean and variance $\mathcal{N}_\circ$. We assume that successful transmission is achieved
when the instantaneous rate is lower than the channel capacity \cite{krikidis2012stability,sadek}.

Let $b$ denote the total number of bits per packet for any node. Since the CR user senses the channel maybe for $\tau_m$ seconds, $m\in\{1,2,\dots,\mathcal{M}\}$, the transmission rate of the CR user is then given by
\begin{equation}
\begin{split}
r_{{\rm s},m}=\frac{b}{T-\tau_m}=\frac{b}{T(1-\frac{\tau_m}{T})}.
\end{split}
\end{equation}

 The outage probability of the link between the CR user and its respective receiver (complement of packet correct reception), when the CR user chooses to sense the channel for $\tau_m$ seconds, is calculated as
\begin{equation}
\begin{split}
P_{{\rm out},{\rm s},m}\!=\!{\rm Pr}\biggr\{\!r_{{\rm s},m}\!>\! \mathcal{C}_{\rm s}\!\biggr\}\!&=\!{\rm Pr}\biggr\{\!r_{{\rm s},m}\!>\!W\log_2\big(\!1\!+\!|h_{\rm s}(t)|^2\gamma_{{\rm s},m}\!\big)\!\biggr\}\\&\!=\!{\rm Pr}\biggr\{|h_{\rm s}(t)|^2\gamma_{{\rm s},m}<2^{\frac{r_{{\rm s},m}}{W}}-1\biggr\}\!\\& =\!1\!-\!\exp\bigg(-{\frac{2^{\frac{r_{{\rm s},m}}{W}}-1}{\sigma^2_{\rm s}\gamma_{{\rm s},m}}}\bigg)\\&=  \!1\!-\!\exp\bigg(-\frac{2^{\frac{b}{WT(1-\frac{\tau_m}{T})}}-1}{\sigma^2_{\rm s}\gamma_{{\rm s},m}}\bigg)
\label{outages}
\end{split}
\end{equation}
\noindent where $r_{{\rm s},m}$ denotes the transmission rate of node `${\rm s}$', $W$ is the channel bandwidth, $\gamma_{{\rm s},m}\!=\!\mathcal{P}_{{\rm s},m}/\mathcal{N}_\circ$ is the received signal-to-noise-ratio (SNR) when the channel gain is unity, and $\mathcal{C}_{\rm s}$ is the capacity of channel `${\rm s}$', and $\sigma_{\rm s}^2$ is the mean of the link gain. If $e$ is the energy used for transmission, then the secondary transmit power, $\mathcal{P}_{{\rm s},m}$, is equal to $e/(T\!-\!\tau_m)$.\footnote{We assume that $e$ is the amount of energy contained in one stored energy packet at the energy queues, i.e., $Q_{\rm pe}$ and $Q_{\rm se}$.} The secondary transmit power proportionally increases with $\tau_m$. Therefore, the delays in transmission due to channel sensing increases the transmitted power and hence the received SNR. However, it also raises the secondary channel outage probability (for proof see the Appendix) and therefore decreases the secondary throughput.

Since the PU accesses the channel at the beginning of the time slot without employing any sensing scheme, the subscript `$m$' is removed from all the primary parameters. The PU transmission rate is $r_{\rm p}\!=\!b/T$. The probability of primary channel outage is then given by
\begin{equation}
\begin{split}
P_{{\rm out},{\rm p}}\!=\!1\!-\!\exp\bigg(-{\frac{2^{\frac{b}{WT}}-1}{\sigma^2_{\rm p}\gamma_{\rm p}}}\bigg)
\end{split}
\end{equation}
\noindent where $\gamma_{\rm p}=\mathcal{P}_{\rm p}/\mathcal{N}_\circ\!=\!e/(T\mathcal{N}_\circ)$ is the received SNR at the primary receiver when the channel gain is unity, and $\mathcal{P}_{\rm p}$ is the power transmitted by the PU.

We assume the use of an energy detector that gathers
a number of samples over a time duration $\tau=\tau_m$, measures their
energy, and then compares the measured energy to a threshold
to make a decision on primary activity \cite{liang2008sensing}. Detection reliability depends on the sensing duration, $\tau_m$. As $\tau_m$ increases, primary detection becomes more
reliable at the expense of reducing the time available for secondary transmission. This is the essence of the sensing-throughput
tradeoff in CR systems \cite{liang2008sensing}.

At the beginning of the time slot, the CR user selects the width of the channel sensing duration from a predefined set $\{\tau_1,\tau_2,\dots,\tau_{\mathcal{M}}\}$. If the PU is sensed to be inactive, the CR user accesses the channel immediately. It should be noted that each sensing duration affects the probability of channel outage between the CR and its respective receiver. The probabilities of false alarm and misdetection are affected as well. Thus, for each sensing duration, $\tau_m\in \{\tau_1,\tau_2,\dots,\tau_{\mathcal{M}}\}$, we obtain three quantities represent the probabilities of channel outage, false alarm and misdetection.

 Let $\alpha^{\rm t}_m=1$ if at time slot $t$ the CR user chooses to sense the channel for $\tau_m$ seconds. At the beginning of each time slot, the CR user selects only one channel sensing duration from the predefined set. Therefore, the optimal duration selection vector, $[\alpha^{\rm t}_1,\alpha^{\rm t}_2,\dots,\alpha^{\rm t}_\mathcal{M}]$, should satisfy the constraint
 \begin{equation}
 \sum_{m=1}^{\mathcal{M}} \alpha^{\rm t}_m=1, \ \forall \ t=0,T,2T,\dots
 \end{equation}
 \section{Service Processes}\label{sec3}
 In this section, we study the service process of each queue in the network. Let ${\rm I}^{\rm t}_{Q_k}$ denote the state of $Q_k$. If $Q_k$ is not empty, ${\rm I}^{\rm t}_{Q_k}=1$, else it is equal to zero. At any time $t$, the PU is active if it has data packets and energy packets. A packet from $Q_{\rm s}$ is served if the CR user decides to sense the channel for $\tau_m$ seconds, the PU is inactive (because either its data or energy queues is empty), the sensor of the CR user does not generate a false alarm, the CR user has energy in its energy queue, and the channel between the CR user and its respective receiver (link `${\rm s}$') is not in outage (ON).\footnote{The channel is ON if it is not in outage, otherwise it is OFF. Recall that the nodes do not have the transmit CSI and therefore the channels' states are unknown.} Mathematically, the service process can be written as:
\begin{equation}
\mathcal{R}^{\rm t}_{\rm s}=  {\rm I}^{\rm t}_{Q_{\rm se}} \big(1-{\rm I}_{Q_{\rm p}}^{\rm t} {\rm I}^{\rm t}_{Q_{\rm pe}}\big) \sum_{m=1}^{\mathcal{M}} \alpha^{\rm t}_m  {\rm I}^{\rm t}_{c_{\rm s},m} \big(1-{\rm I}^{\rm t}_{{\rm FA},m}\big)
\end{equation}
\noindent where ${\rm I}^{\rm t}_{c_{\rm s},m}$ denotes the state of channel `${\rm s}$' when the CR user decides to sense the channel for $\tau_m$. If channel `${\rm s}$' is ON, i.e., it is not in outage, ${\rm I}^{\rm t}_{c_{\rm s},m}=1$, else it is equal to zero.

A packet from the energy queue of the CR user is consumed in either one of the following events. If the PU is active and the CR user misdetects its activity, or if the PU is inactive and the CR user's sensor does not generate a false alarm. Mathematically, the process is given by
\begin{equation}
\begin{split}
\mathcal{R}^{\rm t}_{\rm se}&= {\rm I}^{\rm t}_{Q_{\rm s}}  \sum_{m=1}^{\mathcal{M}} \alpha^{\rm t}_m \bigg[{\rm I}^{\rm t}_{Q_{\rm p}}{\rm I}^{\rm t}_{Q_{\rm pe}}\big(1-{\rm I}^{\rm t}_{D,m}\big)\\& \,\,\,\,\,\,\,\,\,\,\,\,\,\,\,\,\,\,\,\,\,\,\,\,\,\,\,\,\,\,\,\,\,\,\,\,\,\,\,\,\,\,\,\,\,\,\,\,\,\,\,\,\,\,\,\,\,\,\,\,\,\,\,\ +\big(1-{\rm I}^{\rm t}_{Q_{\rm p}}{\rm I}^{\rm t}_{Q_{\rm pe}}\big) (1-{\rm I}^{\rm t}_{{\rm FA},m})\bigg]
\end{split}
\end{equation}
\noindent where ${\rm I}^{\rm t}_{D,m}$ is equal to unity if the CR user at time slot $t$ detects the primary activity correctly and ${\rm I}^{\rm t}_{{\rm FA},m}$ is unity if the CR user's sensor generates a false alarm at time slot $t$.

Since the PU accesses the channel unconditionally at the beginning of the time slot, an energy packet from the primary energy queue is consumed if the primary data queue is nonempty. That is, the service process is given by
\begin{equation}
\mathcal{R}^{\rm t}_{\rm pe}= {\rm I}^{\rm t}_{Q_{\rm p}}.
\end{equation}

 A packet at the head of the primary data queue is served in either one of the following events. When the primary energy queue is nonempty and the channel to its respective receiver (link `${\rm p}$') is ON, 1) if the CR user properly captures the state of primary activity, or 2) if the CR user misdetects the primary activity but either its data queue or energy queue is empty. The service process of the primary data queue, $Q^{\rm t}_{\rm p}$, is given by
\begin{equation}
\begin{split}
    \mathcal{R}^{\rm t}_{\rm p}&=\bigg(1\!-\!{\rm I}^{\rm t}_{Q_{\rm s}}{\rm I}^{\rm t}_{Q_{\rm se}} \sum_{m=1}^{\mathcal{M}} \alpha^{\rm t}_m \big(1-{\rm I}^{\rm t}_{D,m}\big)\bigg){\rm I}^{\rm t}_{c_{\rm p}}{\rm I}^{\rm t}_{Q_{\rm pe}}.
    \end{split}
\end{equation}

Based on the late-arrival model described in Section \ref{sec2}, the evolution of queue $Q_k$ is given by \cite{sadek}
\begin{equation}\label{5}
\begin{split}
    Q^{\rm t+1}_k&=\max\big\{Q^{\rm t}_{k}\!-\!\mathcal{R}^{\rm t}_{k},0\big\}+A^{\rm t}_{k}, \ k={\rm p,s,pe,se}
    \end{split}
\end{equation}
where $A^{\rm t}_{k}\!=\!1$ if a packet is arrived to $Q_{k}$ at time slot $t$ (which occurs with probability $\lambda_k$), else it is equal to zero. The interacting queues render the analysis intractable. Hence, we resort to the concept of dominant systems \cite{rao,sadek,pappas2012optimal,krikidis2012stability}. We consider the case where the PU and CR user send dummy data packets when their data buffers are empty. However, the dummy packets do not contribute the service rates of the data queues, but consume energy packets from the energy queues and cause collisions, i.e., in case of concurrent transmissions. In this case, ${\rm I}^{\rm t}_{Q_{\rm p}}={\rm I}^{\rm t}_{Q_{\rm s}}=1$ at all $t$. Since the PU always has a data packet to send (backlogged), an energy packet is consumed at each time slot. That is,
\begin{equation}
\mathcal{R}^{\rm t}_{\rm pe}= 1.
\label{peeqn}
\end{equation}
The service process of the CR user energy queue is given by
\begin{equation}
\mathcal{R}^{\rm t}_{\rm se}=  \sum_{m=1}^{\mathcal{M}} \alpha^{\rm t}_m \bigg[{\rm I}^{\rm t}_{Q_{\rm pe}}\big(1-{\rm I}^{\rm t}_{D,m}\big)+\big(1-{\rm I}^{\rm t}_{Q_{\rm pe}}\big)(1- {\rm I}^{\rm t}_{{\rm FA},m})\bigg].
\label{seeqn}
\end{equation}

For the data queues, the service processes become
\begin{equation}
\begin{split}
\label{seqn}
\mathcal{R}^{\rm t}_{\rm s}&= {\rm I}^{\rm t}_{Q_{\rm se}} \big(1- {\rm I}^{\rm t}_{Q_{\rm pe}}\big) \sum_{m=1}^{\mathcal{M}} \alpha^{\rm t}_m  {\rm I}^{\rm t}_{c_{\rm s},m} \big(1-{\rm I}^{\rm t}_{{\rm FA},m}\big),\\
    \mathcal{R}^{\rm t}_{\rm p}&=\Big[1\!-\!{\rm I}^{\rm t}_{Q_{\rm se}} \sum_{m=1}^{\mathcal{M}} \alpha^{\rm t}_m \big(1-{\rm I}^{\rm t}_{D,m}\big)\Big]{\rm I}^{\rm t}_{c_{\rm p}}{\rm I}^{\rm t}_{Q_{\rm pe}}.
    \end{split}
\end{equation}
\vspace{-0.4cm}
\section{Stability Analysis and Problem Formulation}\label{sec4}
A fundamental performance measure of a communication network is the stability of its queues, which is rigorously defined in \cite{sadek} for example. If the arrival and service processes are strictly stationary, then we can apply Loynes' theorem to check for stability conditions \cite{loynes1962stability,sadek}. This theorem states that if the arrival process and the service process of a queue are strictly stationary processes, and the average service rate is greater than the average arrival rate of the queue, then the queue is stable, otherwise the queue is unstable. Note that since we use the dominant system approach, the lengths of the data queues in the dominant system always exceed those in the original system provided that all queues are identically initialized. If the dominant system is stable, then the original system must also be stable.
That is, the stability of the dominant system is sufficient for the stability of the original system. Moreover, the achievable service rates in the dominant system would be lower bounds on what can be achieved in the original system.

Let $\mu_k$ denote the mean service rate of $Q_k$, where $\mu_k$ is the expected value of $\mathcal{R}^{\rm t}_{k}$. The mean service rate of the primary energy queue is given by the expectation of (\ref{peeqn}),
\begin{equation}
\begin{split}
\mu_{\rm pe}\!=\! 1.
\end{split}
\end{equation}

Substituting with ${\rm I}^{\rm t}_{Q_{\rm p}}={\rm I}^{\rm t}_{Q_{\rm s}}=1$ into (\ref{seeqn}) and taking the expectation, the mean service rate of the secondary energy queue is given by
\begin{equation}
\begin{split}
 \ \mu_{\rm se}\!=\! \sum_{m=1}^{\mathcal{M}} P_m \bigg(\lambda_{\rm pe}P_{{\rm MD},m}\!+\!\overline{\lambda}_{\rm pe} \overline{P}_{{\rm FA},m}\bigg)
\end{split}
\end{equation}
\noindent where $\overline{x}=1-x$, $P_m$ is the probability that the CR user decides to sense the channel for $\tau_m$ seconds, $ P_{{\rm FA},m}$ is the probability that the CR user's sensor generates false alarm when the CR user chooses the sensing duration to be $\tau_m$ and $P_{{\rm MD},m}\!=\!1\!-\!P_{{\rm D},m}$ is the probability that the CR user misdetects the primary activity. Since both service processes of the energy queues are independent of each other and of the rest of the queues, the probability of having the primary energy queue being empty is given by
\begin{equation}
\begin{split}
{\rm Pr}\big\{Q_{\rm pe}\!=\!0 \big\}&\!=\!1-\frac{\lambda_{\rm pe}}{\mu_{\rm pe}}=1-\lambda_{\rm pe}=\overline{\lambda}_{\rm pe}.
\end{split}
\end{equation}
The probability that the backlogged PU being active is given by $\lambda_{\rm pe}$. For $Q_{\rm se}$, if $\mu_{\rm se}\ge \lambda_{\rm se}$, the Markov chain has a stationary distribution, and the probability of having $Q_{\rm se}\ne0$ is given by
\begin{equation}
\begin{split}
 \mathcal{X}_{\rm se}={\rm Pr}\big\{Q_{\rm se}\ne0 \big\}&\!=\!\frac{\lambda_{\rm se}}{ \sum_{m=1}^{\mathcal{M}} P_m \bigg(\lambda_{\rm pe}P_{{\rm MD},m}\!+\!\overline{\lambda}_{\rm pe} \overline{P}_{{\rm FA},m}\bigg)} .
\end{split}
\end{equation}
If $\mu_{\rm se}< \lambda_{\rm se}$, the energy packets arrival rate is higher than the rate of energy packets consumption and therefore the queue overflows. Consequently, the probability of having $Q_{\rm se}\ne0$ is given by
\begin{equation}
\begin{split}
 {\rm Pr}\big\{Q_{\rm se}\ne0 \big\}&\!=\!1.
\end{split}
\end{equation}
Combining both cases together, we get
\begin{equation}
{\rm Pr}\big\{Q_{\rm se}\ne0 \big\}=  \mathcal{\tilde X}_{\rm se}\!=\!\min\big\{ \mathcal{X}_{\rm se},1\big\}
\end{equation}
where $\min\{.\}$ is the minimum of the values in the argument. For the data queues, the mean service rates are given by
\begin{equation}
\begin{split}
\mu_{\rm p}\!&=\! \lambda_{\rm pe}\overline{P}_{{\rm out},p}\big(1\!-\! \!\mathcal{\tilde X}_{\rm se}{\sum_{m=1}^{\mathcal{M}} P_m P_{{\rm MD},m}} \big),\\
\mu_{\rm s}&=\mathcal{\tilde X}_{\rm se}{\overline{\lambda}_{\rm pe} \sum_{m=1}^{\mathcal{M}} P_m  \overline{P}_{{\rm out},{\rm s},m} \overline{P}_{{\rm FA},m}}.
\label{mus}
\end{split}
\end{equation}
\noindent Using (\ref{mus}) for $\mu_{\rm p}$ and $\mu_{\rm s}$, the CR solves the following optimization problem to obtain the optimal sensing duration probabilities:
 \begin{equation}
\begin{split}
\label{ppoo}
&\underset{P_1,P_2,\dots,P_{\mathcal{M}}}{\max.} \,\,\,\,\,\mu_{\rm s}  \\
 & {\rm s.t.}  \, \, \lambda_{\rm p} \le \mu_{\rm p},\,\, \sum_{m=1}^{\mathcal{M}} P_m=1, \,\ P_m \ge 0 \ \forall m.
    \end{split}
\end{equation}
Based on the relationship among $\mu_{\rm se}$ and $\lambda_{\rm se}$, the optimization problem (\ref{ppoo}) can be split into two optimization subproblems. The first subproblem, when $\lambda_{\rm se}\le \mu_{\rm se}$, is stated as
 \begin{equation}
\begin{split}
&\underset{P_1,P_2,\dots,P_{\mathcal{M}}}{\max.} \,\,\,\,\,\mathcal{X}_{\rm se}{\overline{\lambda}_{\rm pe} \sum_{m=1}^{\mathcal{M}} P_m  \overline{P}_{{\rm out},{\rm s},m} \overline{P}_{{\rm FA},m}}
  \\
 & {\rm s.t.} \,\,\,\,\,\,\ \lambda_{\rm p} \le \lambda_{\rm pe}\overline{P}_{{\rm out},p}\big(1\!-\! \!\mathcal{ X}_{\rm se}{\sum_{m=1}^{\mathcal{M}} P_m P_{{\rm MD},m}} \big),\\ & \,\,\,\,\,\,\,\,\,\,\,\ \lambda_{\rm se}\le \mu_{\rm se},\ \sum_{m=1}^{\mathcal{M}} P_m=1, \,\ P_m \ge 0 \ \forall m.
 \label{opt100}
    \end{split}
\end{equation}
\noindent The optimization problem is a linear-fractional program. It can be readily solved via transformation to a linear program as explained in \cite{boyed}. The second subproblem, when $\lambda_{\rm se}\ge \mu_{\rm se}$, is stated as
 \begin{equation}
\begin{split}
&\underset{P_1,P_2,\dots,P_{\mathcal{M}}}{\max.} \,\,\,\,\, {\overline{\lambda}_{\rm pe} \sum_{m=1}^{\mathcal{M}} P_m  \overline{P}_{{\rm out},{\rm s},m} \overline{P}_{{\rm FA},m}}
  \\
 & {\rm s.t.}  \, \, \lambda_{\rm p} \le \lambda_{\rm pe}\overline{P}_{{\rm out},{\rm p}}\big(1\!-\! \!{\sum_{m=1}^{\mathcal{M}} P_m P_{{\rm MD},m}} \big),\\ & \,\,\,\,\,\,\,\,\,\ \lambda_{\rm se}\ge \mu_{\rm se},\ \sum_{m=1}^{\mathcal{M}} P_m=1, \,\ P_m \ge 0 \ \forall m.
 \label{opt200}
    \end{split}
\end{equation}
\noindent The optimization problem is a linear program. Note that, in this subproblem, $\mu_{\rm s}$ and $\mu_{\rm p}$ are independent of the actual value of $\lambda_{\rm se}$. The optimal solution is taken as the one which yields the highest objective function in (\ref{opt100}) and (\ref{opt200}).

\section{Numerical Results and Conclusions}\label{sec5}
\begin{figure}
   \center
\normalcolor
  \includegraphics[width=0.7\columnwidth]{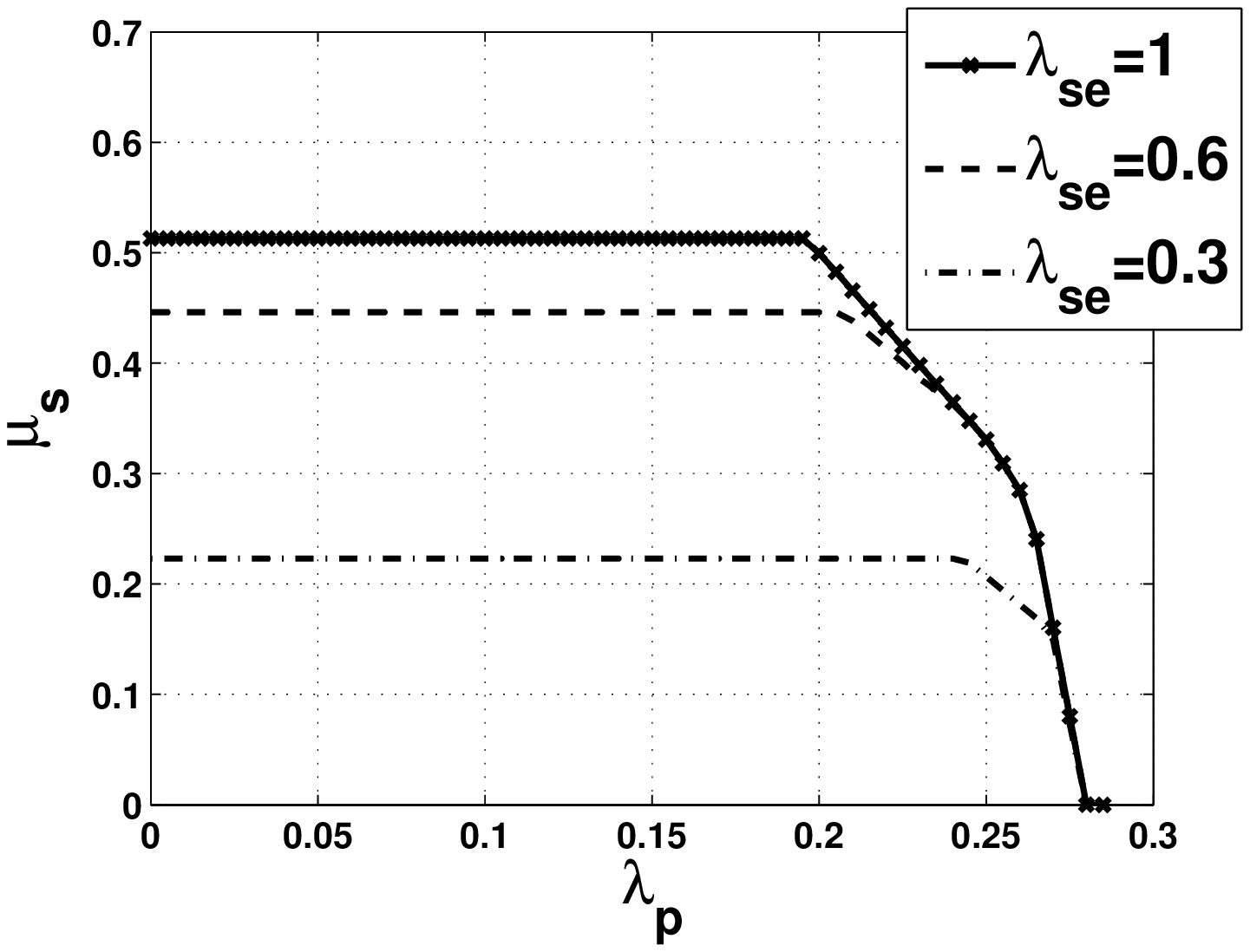}
  \caption{The stability region of the proposed system with different values of the energy queue of the CR user. The figure is generated with $\lambda_{\rm pe}=0.4$ energy packets per time slot.}\label{fig1}
\end{figure}

In this section, we present some numerical results for the optimization problem provided in this paper. Results are generated using primary mean arrival rate $\lambda_{\rm p} \!\in\![0,1]$ packets per time slot. The outage probabilities and channel sensing probabilities are provided in Table \ref{table}, the number of available sensing durations is $\mathcal{M}=10$, and the primary link outage probability is $P_{{\rm out},{\rm p}}=0.3$. Fig. \ref{fig1} presents the maximum secondary stable throughput for each $\lambda_{\rm p}$. The figure shows the impact of increasing the secondary energy queue mean arrival rate. As shown in the figure, the mean service rate of the secondary data queue, $\mu_{\rm s}$, increases with increasing of the mean arrival rate of the secondary energy queue.

\begin{figure}
   \center
\normalcolor
  \includegraphics[width=0.7\columnwidth]{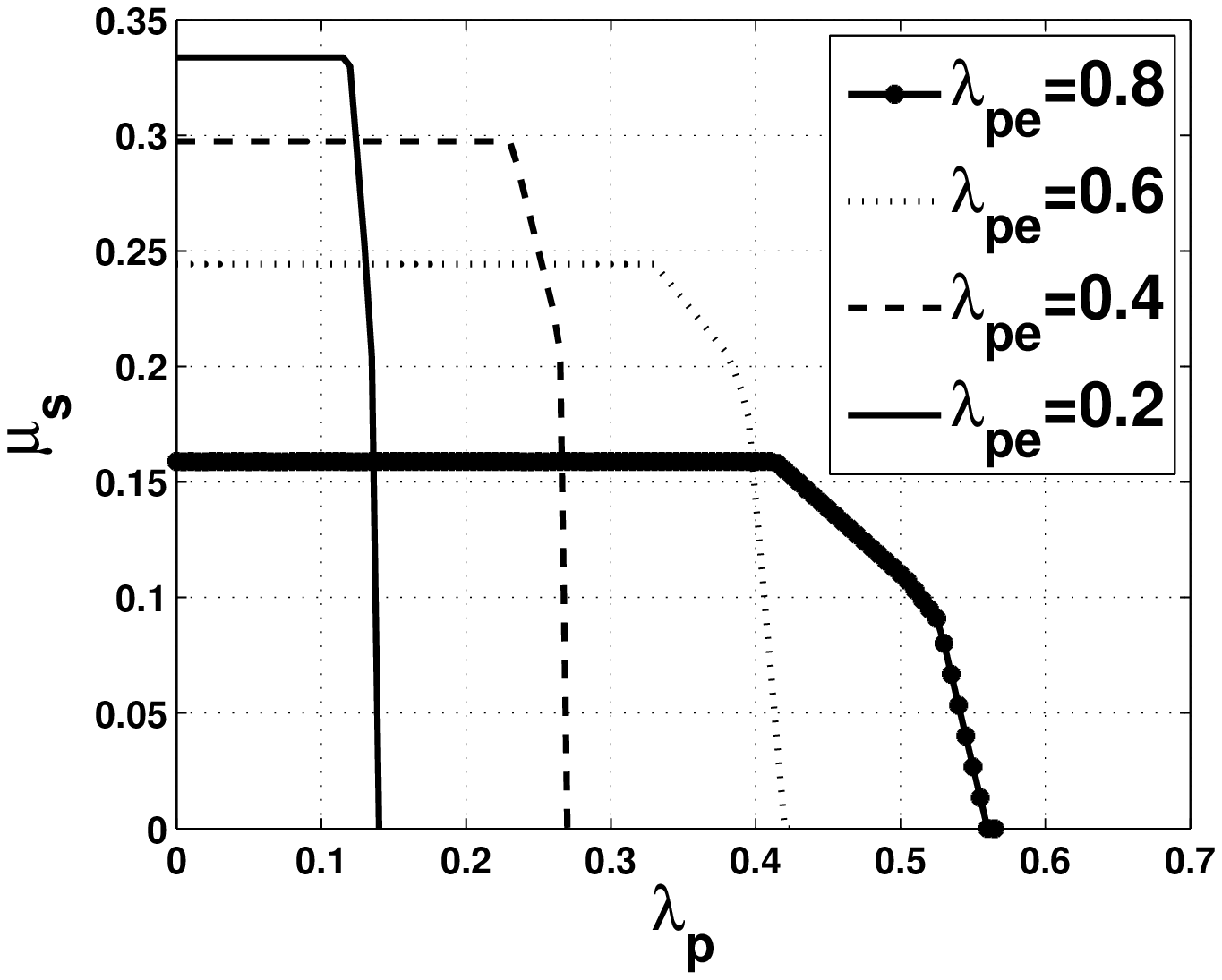}
  \caption{The stability region of the proposed system with different values of the energy queue of the PU. The figure is generated with $\lambda_{\rm se}=0.4$ energy packets per time slot.}\label{fig2}
\end{figure}

\begin{figure}
\normalcolor
\center
  \includegraphics[width=0.7\columnwidth]{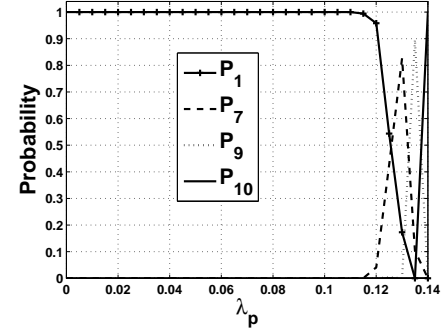}
  \caption{The optimal sensing duration probabilities.}\label{fig3}
\end{figure}

 Fig. \ref{fig2} shows the impact of the mean arrival rate of the PU energy queue. It can be noted that as the mean arrival rate of the primary energy queue increases, the mean service rate of the secondary data queue decreases and the allowable PU mean arrival rate expands. The expansion of the allowable primary arrival occurs because the available energy at the primary energy queue can serve the incoming traffic. Fig. \ref{fig3} presents the optimal values of the sensing duration selection probabilities for different $\lambda_{\rm p}$. The results are generated using Table \ref{table}, $\lambda_{\rm se}=0.4$ packets per time slot and $\lambda_{\rm pe}=0.2$ packets per time slot. It is clear from the figure that as $\lambda_{\rm p}$ increases, the CR user is more likely to select a larger sensing duration in order to enhance the detection reliability of the increasingly active PU.

\begin{figure}
   \center
\normalcolor
  \includegraphics[width=0.7\columnwidth]{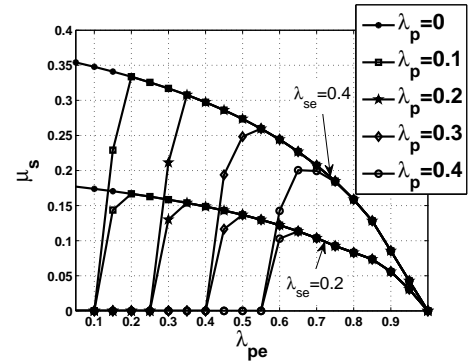}
  \caption{The maximum stable throughput of the CR user versus $\lambda_{\rm pe}$. The figure is generated with different values of $\lambda_p$ and $\lambda_{\rm se}$.}\label{fig4}
\end{figure}

\begin{figure}
   \center
\normalcolor
  \includegraphics[width=0.8\columnwidth]{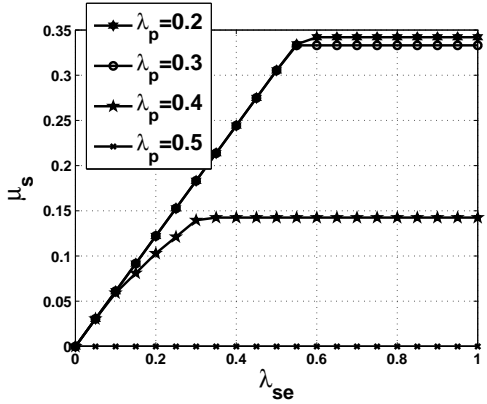}
  \caption{The maximum stable throughput of the CR user versus $\lambda_{\rm se}$. The figure is generated with $\lambda_{\rm pe}=0.6$ energy packets per time slot.}\label{fig5}
\end{figure}

 In Figs. \ref{fig4} and \ref{fig5}, we show the maximum secondary stable throughput versus $\lambda_{\rm pe}$ and $\lambda_{\rm se}$, respectively. It can be noted that the CR user cannot access the channel until the mean arrival rate at the primary queue, $\lambda_{\rm pe}$, achieves certain threshold. This is because the primary queue is unstable below this threshold. As $\lambda_{\rm pe}$ increases, the mean service rate of the secondary data queue decreases. This is because the probability that the backlogged PU is inactive is given by $1\!-\!\lambda_{\rm pe}$ and therefore as $\lambda_{\rm pe}$ increases, the probability that the PU is inactive decreases, and the chances for the CR user to access the channel are reduced. It is also noted that $\lambda_{\rm se}$ controls the maximum value of $\mu_{\rm s}$. More specifically, $\mu_{\rm s}$ at $\lambda_{\rm se}=0.4$ packets per time slot is higher than $\mu_{\rm s}$ at $\lambda_{\rm se}=0.2$ packets per time slot.
  Fig. \ref{fig5} shows the nondecreasing relationship between $\mu_{\rm s}$ and $\lambda_{\rm se}$. As mentioned beneath (\ref{opt200}), when the secondary energy queue is always full, the secondary throughput, $\mu_{\rm s}$, becomes independent of $\lambda_{\rm se}$. This represents the constant parts
in the figure which lead to the interesting fact that the overflow
of the secondary energy queue may not increase the secondary
throughput. This is because the CR user cannot choose lower duration, which corresponds to a lower secondary channel outage probability and a higher misdetection probability, to avoid the increase of collisions with the PU caused by sensing errors. When $\lambda_{\rm p}=0.5$, the PU is unstable. Hence, the CR user cannot access the channel.

\begin{figure}
   \center
\normalcolor
  \includegraphics[width=0.8\columnwidth]{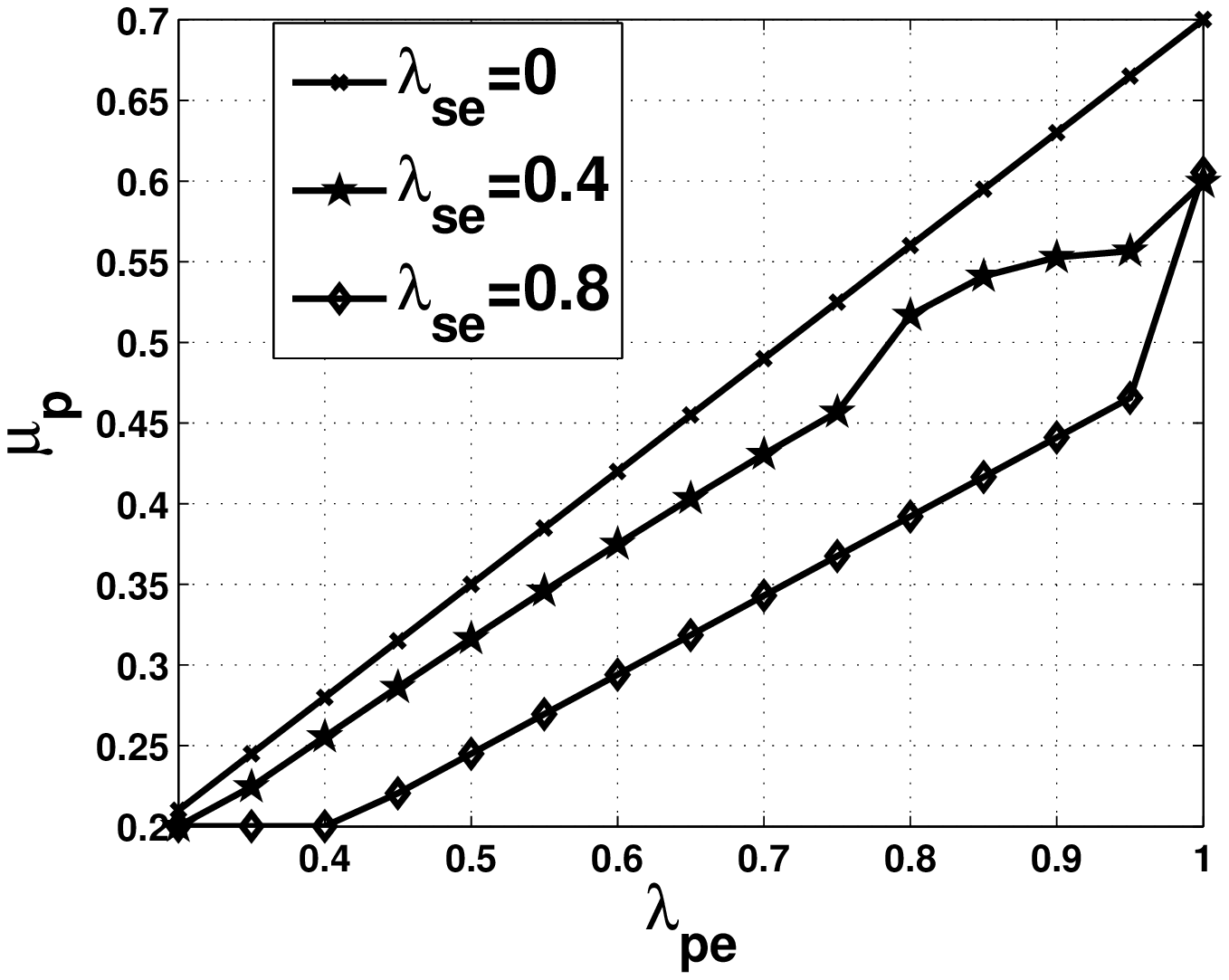}
  \caption{The primary mean service rate versus $\lambda_{\rm pe}$. The figure is generated with $\lambda_{\rm p}=0.2$ packets per time slot.}\label{fig6}
\end{figure}

\begin{figure}
   \center
\normalcolor
  \includegraphics[width=0.8\columnwidth]{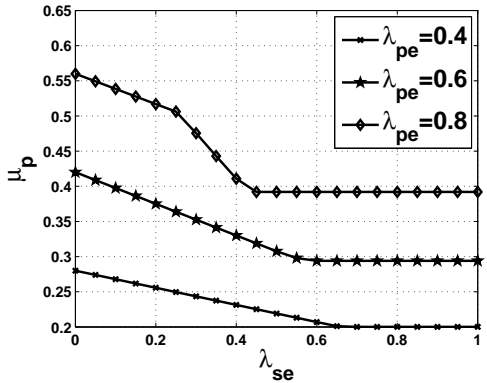}
  \caption{The primary mean service rate versus $\lambda_{\rm se}$. The figure is generated with $\lambda_{\rm p}=0.2$ packets per time slot.}\label{fig7}
\end{figure}

 In Figs. \ref{fig6} and \ref{fig7}, we show the primary stable throughput versus $\lambda_{\rm pe}$ and $\lambda_{\rm se}$, respectively. The service rate of the primary data queue, $\mu_{\rm p}$, is increasing with $\lambda_{\rm pe}$ and nondecreasing with $\lambda_{\rm se}$. This is because as the secondary energy arrival increases, the CR user will be able to access the channel for more slots, thereby causing more collisions with the PU due to sensing errors. The constant part occurs because the secondary energy queue is always full (i.e., ${\rm Pr}\{Q_{\rm se}\ne0\}=1$) which results in independency of $\mu_{\rm p}$ and $\mu_{\rm s}$ on $\lambda_{\rm se}$ as mentioned earlier beneath the second subproblem. Note that $\mu_{\rm p}$ must not be lower than $\lambda_{\rm p}$ to ensure primary queue stability. Specifically, for the given parameters, $\mu_{\rm p}\ge \lambda_{\rm p}=0.2$ packets/time slot.

 In the paper, we have investigated the maximum stable-throughput of an energy harvesting CR user in presence of an energy harvesting PU. The CR user optimally selects one of the available sensing durations from a predefined set to maximize its throughput. One possible extension is to consider more general channel
models which allow simultaneous transmissions and multipacket
reception.
    \begin{table*}
    \centering
\begin{tabular}{|@{}c@{}|c|c|c|c|c|c|c|c|c|c|}
    \hline\hline & $\tau_1$&$\tau_2$&$\tau_3$&$\tau_4$&$\tau_5$&$\tau_6$&$\tau_7$&$\tau_8$&$\tau_9$&$\tau_{10}$
    \\[5pt]\hline  $P_{D,m}$ &$0.7$&$0.75$&$0.78$&$0.8$&$0.85$&$0.88$&$0.9$&$0.92$&$0.94$&$0.95$ \\[5pt]\hline
     $P_{{\rm FA},m}$ &$0.05$ &$0.06$ &$0.08$ &$0.082$ &$0.085$ &$0.088$ &$0.1$ &$0.11$ &$0.12$ &$0.125$ \\[5pt]\hline
      $P_{{\rm out},{\rm s},m}$& $0.1$ &$0.2$ &$0.25$ &$0.3$ &$0.35$ &$0.38$ &$0.4$ &$0.46$ &$0.49$ &$0.6$ \\[5pt]\hline
\end{tabular}
\caption{The values of outage probability, false-alarm probability, and misdetection probability corresponding to $[\tau_1,\tau_2,\dots,\tau_{10}]$}
\label{table}
\end{table*}
\section*{Appendix}
\begin{proof}
We prove here that $P_{{\rm out},{\rm s},\nu} > P_{{\rm out},{\rm s},n}$ where $\tau_\nu>\tau_n$. As a function of $\tau_m$, $P_{{\rm out},{\rm s},m}$ is given by (\ref{outages}). Let $z_m=1-\frac{\tau_m}{T}$ where
$z_m \in [0,1]$. Assuming that the energy unit used per time slot is $e$, the transmit power is $\frac{e}{T(1-\frac{\tau_m}{T})}=\frac{e}{Tz_m}$. This means that the received SNR $\gamma_{m,s}$ is inversely proportional to $z_m$. The exponent in (\ref{outages}) is thus proportional to $\boldsymbol{g(z_m)=z_m (\exp(\frac{a}{z_m})-1)}$ where $a =\frac{b \ln 2} {WT} > 0$. Differentiating $g(z_m)$ with respect to $z_m$, the derivative is given by
\begin{equation}
\begin{split}
\label{vvcc}
 \tilde g(z_m)=-1+\Big[1-\frac{a}{z_m}\Big] \exp(\frac{a}{z_m})&
\end{split}
\end{equation}
where $a/z_m >0$. The first term in (\ref{vvcc}) is a negative constant. To prove the negatively of $\tilde g(z_m)$, the second term should be negative. If $\frac{a}{z_m}>1$, then $\big[1-\frac{a}{z_m}\big]<0$, $\exp(\frac{a}{z_m})>1$, and the derivative is non-positive, i.e., $\tilde g(z_m) \le 0$.
If $\frac{a}{z_m}\le1$, then $[1-\frac{a}{z_m}]\le1$, $\exp(\frac{a}{z_m})\le1$, and the derivative is non-positive, i.e., $ \tilde g(z_m)<0$.
Therefore, the derivative is always negative. Since $z_m=1-\frac{\tau_m}{T}$, function $g(z_m)$ increases with $\tau_m$. This means that $\overline{P}_{{\rm out},m,s}$ decreases with $\tau_m$ and its maximum value occurs when the transmission starts at the beginning of the time slot, i.e., $\tau_m=0$. This proves that $P_{{\rm out},\nu,s} > P_{{\rm out},n,s}$ where $\tau_\nu>\tau_n$.
\end{proof}
\bibliographystyle{IEEEtran}
\bibliography{IEEEabrv,nw}

\begin{thebibliography}{10}
\providecommand{\url}[1]{#1}
\csname url@samestyle\endcsname
\providecommand{\newblock}{\relax}
\providecommand{\bibinfo}[2]{#2}
\providecommand{\BIBentrySTDinterwordspacing}{\spaceskip=0pt\relax}
\providecommand{\BIBentryALTinterwordstretchfactor}{4}
\providecommand{\BIBentryALTinterwordspacing}{\spaceskip=\fontdimen2\font plus
\BIBentryALTinterwordstretchfactor\fontdimen3\font minus
  \fontdimen4\font\relax}
\providecommand{\BIBforeignlanguage}[2]{{%
\expandafter\ifx\csname l@#1\endcsname\relax
\typeout{** WARNING: IEEEtran.bst: No hyphenation pattern has been}%
\typeout{** loaded for the language `#1'. Using the pattern for}%
\typeout{** the default language instead.}%
\else
\language=\csname l@#1\endcsname
\fi
#2}}
\providecommand{\BIBdecl}{\relax}
\BIBdecl

\bibitem{lei2009generic}
J.~Lei, R.~Yates, and L.~Greenstein, ``A generic model for optimizing
  single-hop transmission policy of replenishable sensors,'' \emph{IEEE Trans.
  Wireless Commun.}, vol.~8, no.~2, pp. 547--551, Feb. 2009.

\bibitem{sharma2010optimal}
V.~Sharma, U.~Mukherji, V.~Joseph, and S.~Gupta, ``Optimal energy management
  policies for energy harvesting sensor nodes,'' \emph{IEEE Trans. Wireless
  Commun.}, vol.~9, no.~4, pp. 1326--1336, Apr. 2010.

\bibitem{ho2010optimal}
C.~Ho and R.~Zhang, ``Optimal energy allocation for wireless communications
  powered by energy harvesters,'' in \emph{IEEE ISIT}, Jun. 2010, pp.
  2368--2372.

\bibitem{hoang2009opportunistic}
A.~Hoang, Y.~Liang, D.~Wong, Y.~Zeng, and R.~Zhang, ``Opportunistic spectrum
  access for energy-constrained cognitive radios,'' \emph{IEEE Trans. Wireless
  Commun.}, vol.~8, no.~3, pp. 1206--1211, Mar. 2009.

\bibitem{park2011energy}
S.~Park, S.~Lee, B.~Kim, D.~Hong, and J.~Lee, ``Energy-efficient opportunistic
  spectrum access in cognitive radio networks with energy harvesting,'' in
  \emph{Proc. of the 4th Int. Conf. on CogART}.\hskip 1em plus 0.5em minus
  0.4em\relax ACM, 2011, pp. 1--5.

\bibitem{krikidis2012stability}
I.~Krikidis, T.~Charalambous, and J.~Thompson, ``Stability analysis and power
  optimization for energy harvesting cooperative networks,'' \emph{IEEE Signal
  Process. Lett.}, vol.~19, no.~1, pp. 20--23, Jan. 2012.

\bibitem{pappas2012optimal}
N.~Pappas, J.~Jeon, A.~Ephremides, and A.~Traganitis, ``Optimal utilization of
  a cognitive shared channel with a rechargeable primary source node,'' in
  \emph{JCN}, vol.~14, no.~2, Apr. 2012, pp. 162--168.

\bibitem{Sultan}
A.~Sultan, ``Sensing and transmit energy optimization for an energy harvesting
  cognitive radio,'' vol.~1, no.~5, pp. 500--503, Oct. 2012.

\bibitem{Sult1210:Optimal}
A.~{El Shafie} and A.~Sultan, ``Optimal random access and random spectrum
  sensing for an energy harvesting cognitive radio,'' in \emph{WiMob}, Oct.
  2012, pp. 403--410.

\bibitem{sadek}
A.~Sadek, K.~Liu, and A.~Ephremides, ``Cognitive multiple access via
  cooperation: protocol design and performance analysis,'' \emph{IEEE Trans.
  Inf. Theory.}, vol.~53, no.~10, pp. 3677--3696, Oct. 2007.

\bibitem{medepally2009implications}
B.~Medepally, N.~Mehta, and C.~Murthy, ``Implications of energy profile and
  storage on energy harvesting sensor link performance,'' in \emph{IEEE
  GLOBECOM}, Dec. 2009, pp. 1--6.

\bibitem{michelusi2012optimal}
N.~Michelusi, K.~Stamatiou, and M.~Zorzi, ``On optimal transmission policies
  for energy harvesting devices,'' in \emph{IEEE ITA}, Feb. 2012, pp. 249--254.

\bibitem{liang2008sensing}
Y.~Liang, Y.~Zeng, E.~Peh, and A.~Hoang, ``Sensing-throughput tradeoff for
  cognitive radio networks,'' \emph{IEEE Trans. Wireless Commun.}, vol.~7,
  no.~4, pp. 1326--1337, Apr. 2008.

\bibitem{rao}
R.~Rao and A.~Ephremides, ``On the stability of interacting queues in a
  multiple-access system,'' \emph{IEEE Trans. Inf. Theory.}, vol.~34, no.~5,
  pp. 918--930, Sept. 1988.

\bibitem{loynes1962stability}
R.~Loynes, ``The stability of a queue with non-independent inter-arrival and
  service times,'' in \emph{Proc. Cambridge Philos. Soc}, vol.~58, no.~3.\hskip
  1em plus 0.5em minus 0.4em\relax Cambridge Univ. Press, July 1962, pp.
  497--520.

\bibitem{boyed}
S.~Boyd and L.~Vandenberghe, ``Convex optimization problems,'' in \emph{Convex
  Optimization}.\hskip 1em plus 0.5em minus 0.4em\relax Cambridge Univ. Press,
  2004, ch.~4, p. 151.

\end{thebibliography}
\end{document}